\documentclass[twocolumn,twocolappendix]{emulateapj}
\usepackage{natbib}
\usepackage{graphics,graphicx,subfigure}

\begin{document}
\title{Grain size constraints on HL Tau with polarization signature}

\author{Akimasa Kataoka \altaffilmark{1,2}, Takayuki Muto\altaffilmark{3}, Munetake Momose \altaffilmark{4}, Takashi Tsukagoshi \altaffilmark{4}, Cornelis P Dullemond\altaffilmark{1}}
\affil{\altaffilmark{1}Zentrum f\"ur Astronomie der Universit\"at Heidelberg, Institut f\"ur Theoretische Astrophysik, Albert-Ueberle-Str. 2, 69120 Heidelberg, Germany}
\email{kataoka@uni-heidelberg.de}
\affil{\altaffilmark{2}National Astronomical Observatory of Japan, Mitaka, Tokyo 181-8588, Japan}
\affil{\altaffilmark{3}Division of Liberal Arts, Kogakuin University, 1-24-2 Nishi-Shinjuku, Shinjuku-ku, Tokyo 163-8677, Japan}
\affil{\altaffilmark{4}College of Science, Ibaraki University, 2-1-1 Bunkyo, Mito, Ibaraki 310-8512, Japan}

\keywords{dust, polarization, protoplanetary disks}

\begin{abstract}
The millimeter-wave polarization of the protoplanetary disk around HL Tau has been interpreted as the emission from elongated dust grains aligned with the magnetic field in the disk.
However, the self-scattering of thermal dust emission may also explain the observed millimeter-wave polarization.
In this paper, we report a modeling of the millimeter-wave polarization of the HL Tau disk with the self-polarization.
Dust grains are assumed to be spherical and to have a power-law size distribution.
We change the maximum grain size with a fixed dust composition in a fixed disk model to find the grain size to reproduce the observed signature.
We find that the direction of the polarization vectors and the polarization degree can be explained with the self-scattering.
Moreover, the polarization degree can be explained only if the maximum grain size is $\sim 150 {\rm~\mu m}$.
The obtained grain size from the polarization is different from a size of millimeter or larger that has been previously expected from the spectral index of the dust opacity coefficient if the emission is optically thin. 
We discuss that porous dust aggregates may solve the inconsistency of the maximum grain size between the two constraints.
\end{abstract}

\flushbottom
\maketitle

\section{Introduction}

Protoplanetary disks are believed to be the birthplace of planets.
Due to the high spatial dust density, submicron-sized dust grains coagulate to form larger bodies, and ultimately form planets in the disks \citep[e.g.,][]{Weidenschilling77, Hayashi81}.
Thus, observational constraints on the grain size in protoplanetary disks are essential to directly investigate the ongoing planet formation.

The grain size in protoplanetary disks has been constrained with the spectral index of the dust opacity coefficient at millimeter wavelengths \citep[e.g.,][]{Beckwith90, BeckwithSargent91,MiyakeNakagawa93, AndrewsWilliams05, Isella09, Ricci10a, Ricci10b, vanderMarel13}.
The opacity index at millimeter wavelengths in protoplanetary disks has been shown to be as low as $0-1$, which indicates that the grain size in protoplanetary disks is of the order of millimeter, or larger \citep[e.g.,][]{Draine06}.
In addition, even at some of the envelopes of protostars, the opacity index is indicated to be as low as the later stage of protoplanetary disks, which suggests that the dust grains are grown to millimeter in size even in the early phase of the protoplanetary disks \citep{Miotello14} although the modeling contains large uncertainties.
However, the low opacity index can also be explained with optically thick disks \citep[e.g.,][]{Ricci12b}, irregularly shaped grains \citep[e.g.,][]{Min05, Min07}, or different chemical composition of the grains \citep[e.g.,][]{Pollack94, Mennella98, Jones12b}.
Therefore, there are still large uncertainties of constraints on the grain size in protoplanetary disks.

Recently, an independent method to constrain the grain size in protoplanetary disks has been proposed.
\citealt{Kataoka15} proposed that millimeter-wave emission is partially polarized due to the self-polarization if the following two condition are satisfied.
The first condition is that dust grains have the size comparable to the wavelengths to have the scattering efficiency large enough to produce the scattered emission.
The second is that the thermal dust emission has anisotropic distributions so that the thermal dust emission is scattered by dust grains themselves and to show the residual polarization in total due to the anisotropic distribution of the incoming fluxes.
Therefore, if the polarization due to the self-scattering of dust thermal emission is detected, it would be the evidence of the existence of dust grains which have the comparable size to the wavelength.
In this paper, we report the application of this method to the protoplanetary disk around HL Tau.

HL Tau is a protostar 140 pc away from the sun \citep{Rebull04}.
This young star is surrounded by an envelope and producing jets and outflows, which indicates a highly active star formation \citep[e.g.,][]{Hayashi93}.
The millimeter emission from the envelope and the disk has been intensively investigated with interferometers \citep{Beckwith90, Mundy96, Looney00, Wilner96, Greaves08, Kwon11}.
Furthermore, Atacama Large Millimeter/submillimeter Array (ALMA) has revealed that the circumstellar disk around HL Tau has multiple rings at sub-millimeter wavelengths, which might be a signature of multiple planets \citep{Partnership15}.
In addition, spatially resolved polarized emission of the disk at sub-mm wavelengths has been detected with Combined Array for Millimeter-wave Astronomy (CARMA) and Submillimeter Array (SMA).
The net polarization degree is $0.89 \%$ in average with CARMA, and $0.86 \pm 0.4 \%$ with SMA.
The polarization vectors are directed from north-east to south-west.
In this paper, we construct a disk model which reasonably reproduces the ALMA observations and perform radiative transfer calculations to investigate the polarization due to the self-scattering of thermal dust emission.

The millimeter-wave polarization observations toward star-forming regions and protoplanetary disks have been interpreted as an indicator of the morphology of the magnetic field \citep{Girart06, Girart09, Hull13, Hull14, Rao14, Stephens14, Segura-Cox15}.
The elongated dust grains have a preferential direction perpendicular to the magnetic field, which results in the polarization of the thermal dust emission \citep{DavisGreenstein51}.
In addition, radiative torques help dust grains to be aligned with the magnetic field \citep{DraineWeingartner97,ChoLazarian05,ChoLazarian07, HoangLazarian08,HoangLazarian09a,HoangLazarian09b}.

This paper does not intend to exclude the possibility that the polarized emission of HL Tau disk is caused by the grain alignment with the disk magnetic field but investigates the self-polarization as an alternative explanation for the polarization observed with CARMA and SMA.
We also discuss possible methods to distinguish the mechanisms between the grain alignment and the self-polarization.

Note that, during the preparation of the manuscript, \citealt{Yang15} independently found that the interpretation of the millimeter-wave polarization of HL Tau due to dust scattering.

\section{Method}

We construct an axisymmetric dust disk model with several gaps with a smooth temperature distribution which reasonably reproduces the sub-mm continuum image at $\lambda = 1.3 {\rm~mm}$ \citep{Partnership15}.
This model is not a unique solution of the density and temperature of the HL Tau disk. 
However, this model would be enough to investigate the grain size with polarization signature as long as the millimeter-wave continuum is reasonably reproduced.
The polarization fraction is determined by the combination of the grain size and the anisotropy of the thermal emission.
The difference in the temperature and density distributions affects the anisotropy of the radiation field, but the dependence of the polarization on the grain size is thought to be stronger.
We discuss this point in Section \ref{sec:discussion}, but the detailed parameter studies of the density and temperature remain to be a future work.

First, we fix the temperature profile with the smooth power-law distribution as
\begin{equation}
T=T_{0}\left(\frac{R}{1{\rm AU}}\right)^{-q_{\rm t}}.
\label{eq:temperature}
\end{equation}
Then, we add the additional gaps on the density profile to reproduce the observation along the major axis of the disk observed with the continuum observed with the ALMA Band 6.
\begin{eqnarray}
\label{eq:tau}
\tau = \tau_{0}\left(\frac{R}{1{\rm AU}}\right)^{-p}\exp{\left(-\left(\frac{R}{R_{\rm exp}}\right)^{s_{\rm exp}}\right)}\\ \nonumber
\times \sum_{i} \left(1-\exp\left(-\frac{1}{2}\left(\frac{R-r_{d,i}}{w_{d,i}}\right)^{2}\right)/f_{d,i}\right).
\end{eqnarray}
The parameters are summarized in Tables \ref{table:disk} and \ref{table:gap}.

\begin{figure}[ht]
\centering
\includegraphics[width=\linewidth]{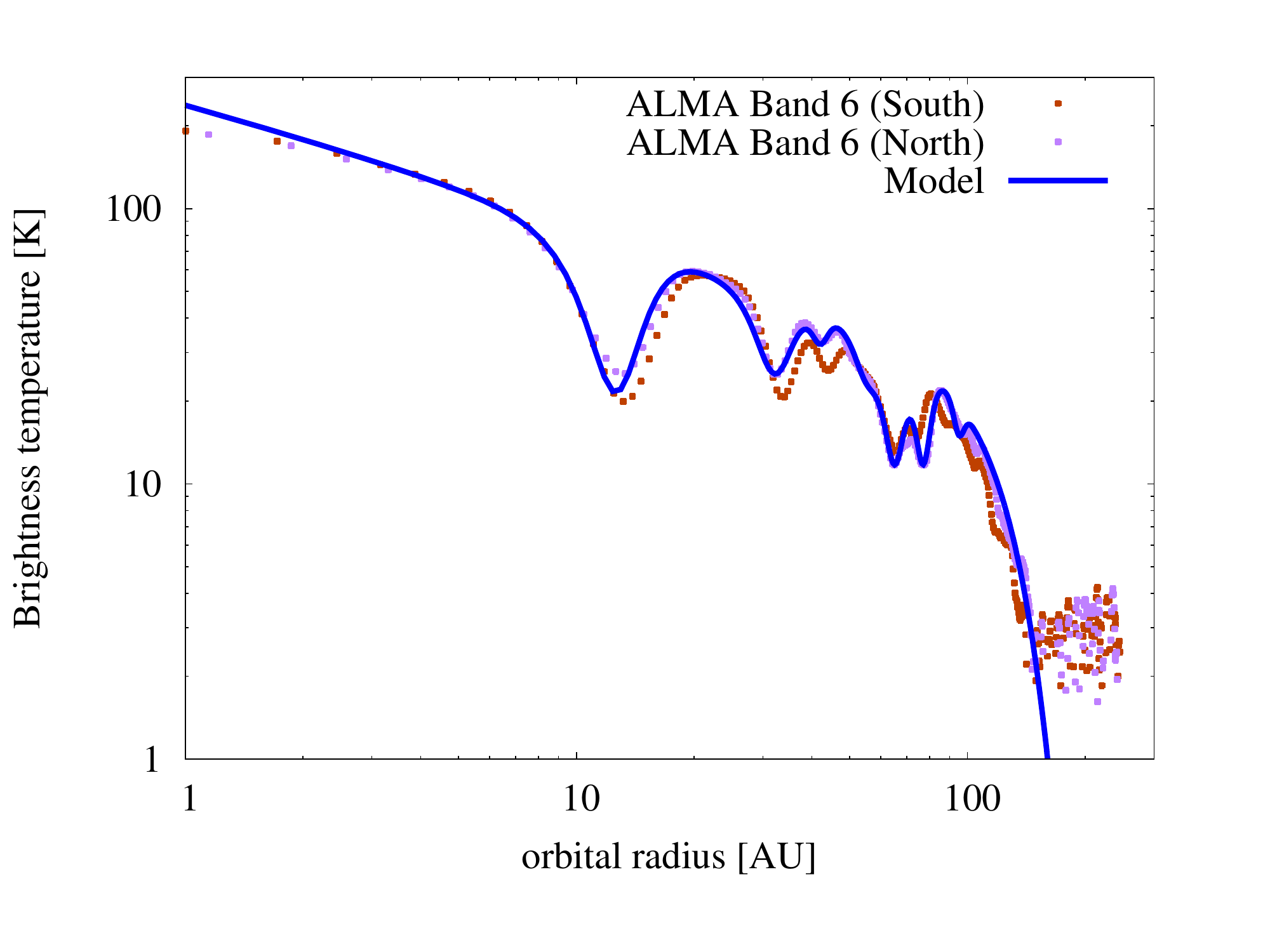}
\caption{
The data points show the observed intensity profile of Band 6 of ALMA along the major axis in brightness temperature.
The solids line shows the brightness temperature of the disk model adopted in this paper.
}
\label{fig:model}
\end{figure}

Figure \ref{fig:model} shows the brightness temperature derived from the Band 6 data on the major axis illustrated with the model brightness temperature, which is calculated as
\begin{equation}
I_{\nu}(R)=\left(1-\exp(-\tau(R)\right)B_{\nu}(T(R)).
\end{equation}
Then, we model the disk with dust surface density of $\Sigma_{\rm d}=\tau(R)/\kappa_{\rm abs}(a_{\rm max}, \lambda=1.3 {\rm~mm})$, where $\kappa_{\rm abs}(a_{\rm max}, \lambda)$ is the absorption opacity including a size distribution of dust grains.

In calculating the opacity, the dust grains are assumed to be spherical and have a power-law size distribution with a power of $q=-3.5$ \citep{Mathis77} with the maximum grain size $a_{\rm max}$.
We take this maximum grain size as the representative grain size in the following discussion.
The opacity is calculated with Mie theory. 
The composition is assumed to be the mixture of silicate, organics, and water ice \citep{Kataoka14,Pollack94}.
We use the refractive index of astronomical silicate \citep{WeingartnerDraine01}, organics \citep{Pollack94}, and water ice \citep{Warren84} and calculate the mixture of them with the effective medium theory with the Maxwell-Garnett rule \citep[e.g.,][]{BohrenHuffman83, MiyakeNakagawa93}.
A different abundance may lead to the different absolute value of polarization degree, which should be investigated in future studies.
The adopted value for the fiducial run is $\kappa_{\rm abs}(a_{\rm max}=150 {\rm~\mu m}, \lambda=1.3 {\rm~mm})=0.24 {\rm~cm^{2}~g^{-1}}$.
In the fiducial case, therefore, the dust surface density $\Sigma_{\rm d}$ has a profile of
\begin{eqnarray}
\Sigma_{\rm d} &=& 8.3 {\rm~g~cm^{-2}} \left(\frac{R}{1{\rm~AU}}\right)^{-p} \exp{\left(-\left(\frac{R}{R_{\rm exp}}\right)^{s_{\rm exp}}\right)}\\ \nonumber
 &&\times \sum_{i} \left(1-\exp\left(-\frac{1}{2}\left(\frac{R-r_{d,i}}{w_{d,i}}\right)^{2}\right)/f_{d,i}\right).
\end{eqnarray}

\begin{deluxetable}{cclcccc}
\tabletypesize{\scriptsize}
\tablecaption{The disk parameters}
\tablewidth{0pt}
\tablehead{
parameters & values
}
\startdata
$T_0$ & 280 K\\
$q_{\rm t}$ & 0.3\\
$\tau_0$&2.0\\
$p$ & 0.3\\
$s_{\rm exp}$ & 4\\
$R_{\rm exp}$ & 120 AU
\enddata
\label{table:disk}
\end{deluxetable}

\begin{deluxetable}{cclcccc}
\tabletypesize{\scriptsize}
\tablecaption{The gap parameters}
\tablewidth{0pt}
\tablehead{
gap number & $r_{d,i}$ & $w_{d,i}$  & $f_{d,i}$ 
}
\startdata
1 & 12.5 & 3.0 & 1.25\\
2 & 32.0 & 4.0 & 1.0 \\
3 & 42.0 & 2.0 & 4.0 \\
4 & 55.0 & 4.0 & 3.0 \\
5 & 65.0 & 4.0 & 1.4 \\
6 & 77.0 & 6.0 & 1.5 \\
7 & 95.0 & 5.0 & 2.0 
\enddata
\label{table:gap}
\end{deluxetable}

We perform the radiative transfer simulations with a public code RADMC-3D to obtain the dust continuum, the polarized intensity, and the polarization degree.
To obtain the vertical density distribution, we assume the Gaussian density distribution with a dust scale height $h_{\rm d}$ such that $\rho_{\rm d}=\Sigma_{\rm d}/(\sqrt{2\pi }h_{\rm d}) \exp(-z^2/2h_{\rm d}^2)$.
To reproduce the geometrically flat disk observed with ALMA, we set $h_{\rm d}=h_{\rm g}/f_{\rm settle}$, where $h_{\rm g}$ is the gas pressure scale height and $f_{\rm settle}=10$ (see Appendix).
Here, we do not use the thermal Monte Carlo simulations to determine the temperature but use the simple power-law temperature model described above.
We assume that the distance to HL Tau is 140 pc, so 1 arcsec corresponds to 140 AU in the figures.
The inclination is assumed to be $40^\circ$.

Note again that this modeling is not a unique solution to reproduce the emission of the HL Tau disk.
However, the main goal of this paper is to constrain the grain size from the polarization observations. 
As we discuss in the following sections, the polarization degree is mainly determined by the combination of the observed wavelengths and the grain size, which does not so much depend on the detailed modeling of temperature and the surface density.

\section{Results}

\begin{figure*}[ht]
\centering
 \subfigure{
 \hspace{-100pt}
\includegraphics[width=100mm]{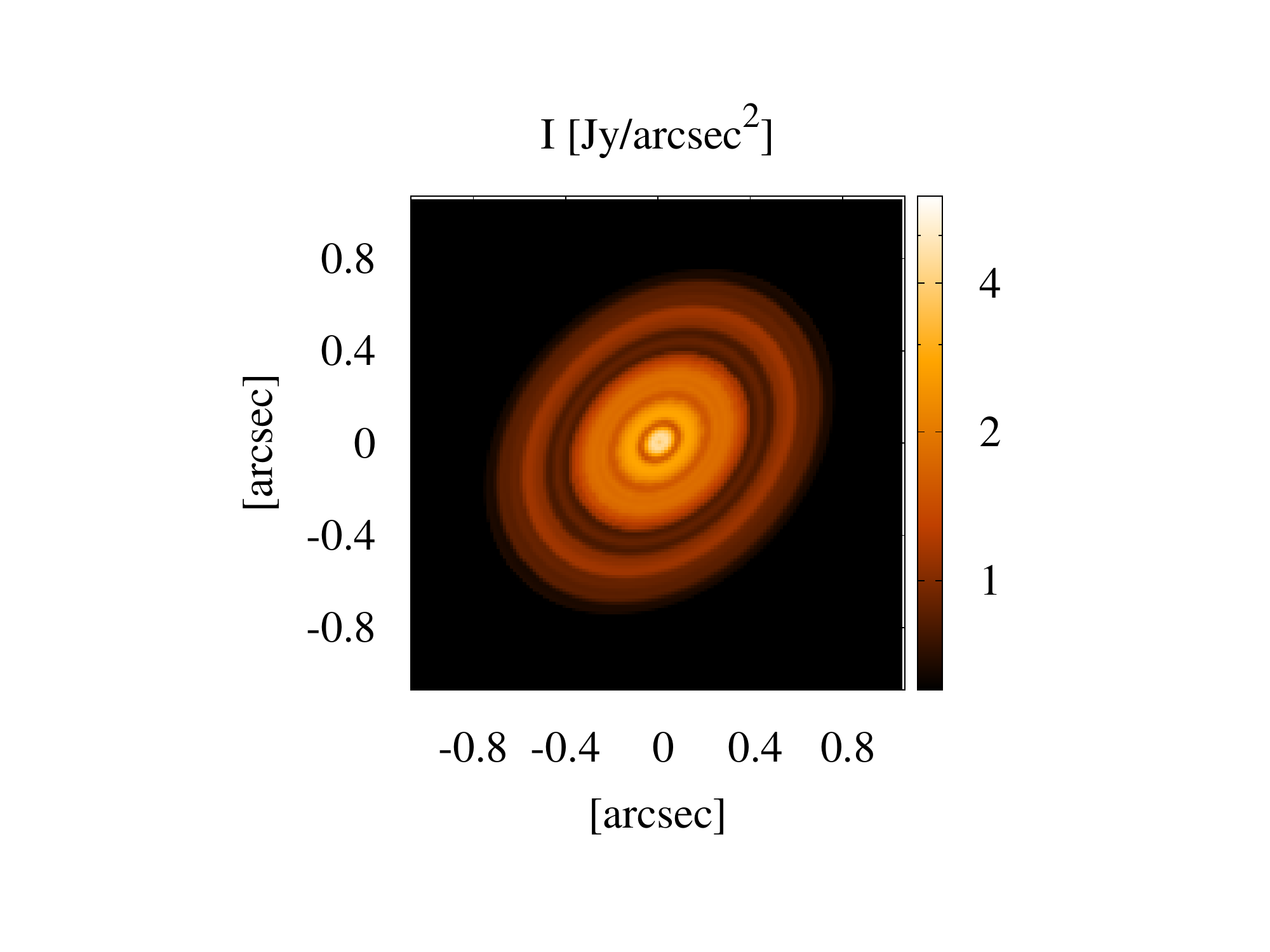}
}
 \subfigure{
 \hspace{-120pt}
\includegraphics[width=100mm]{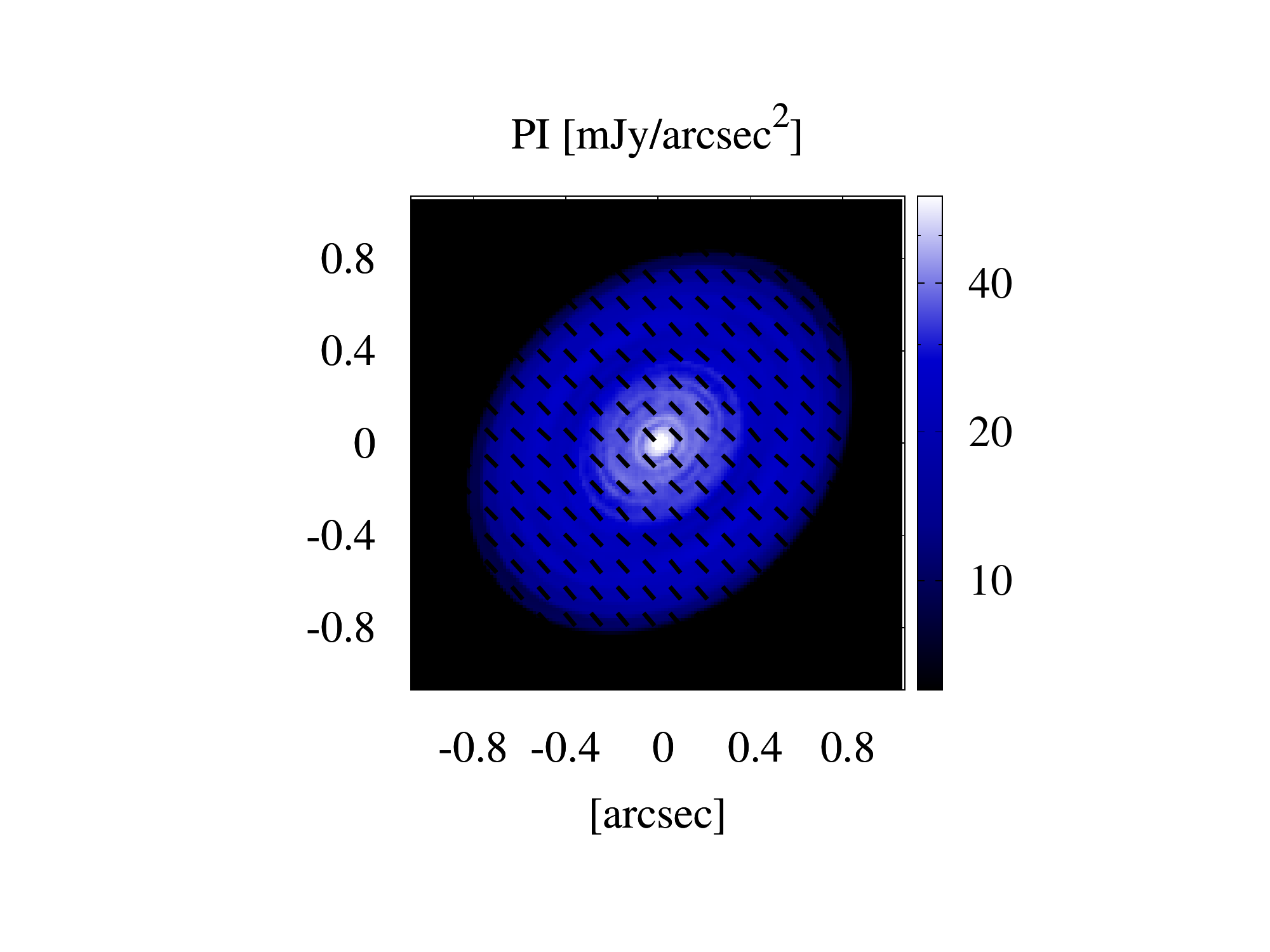}
}
 \subfigure{
  \hspace{-120pt}
\includegraphics[width=100mm]{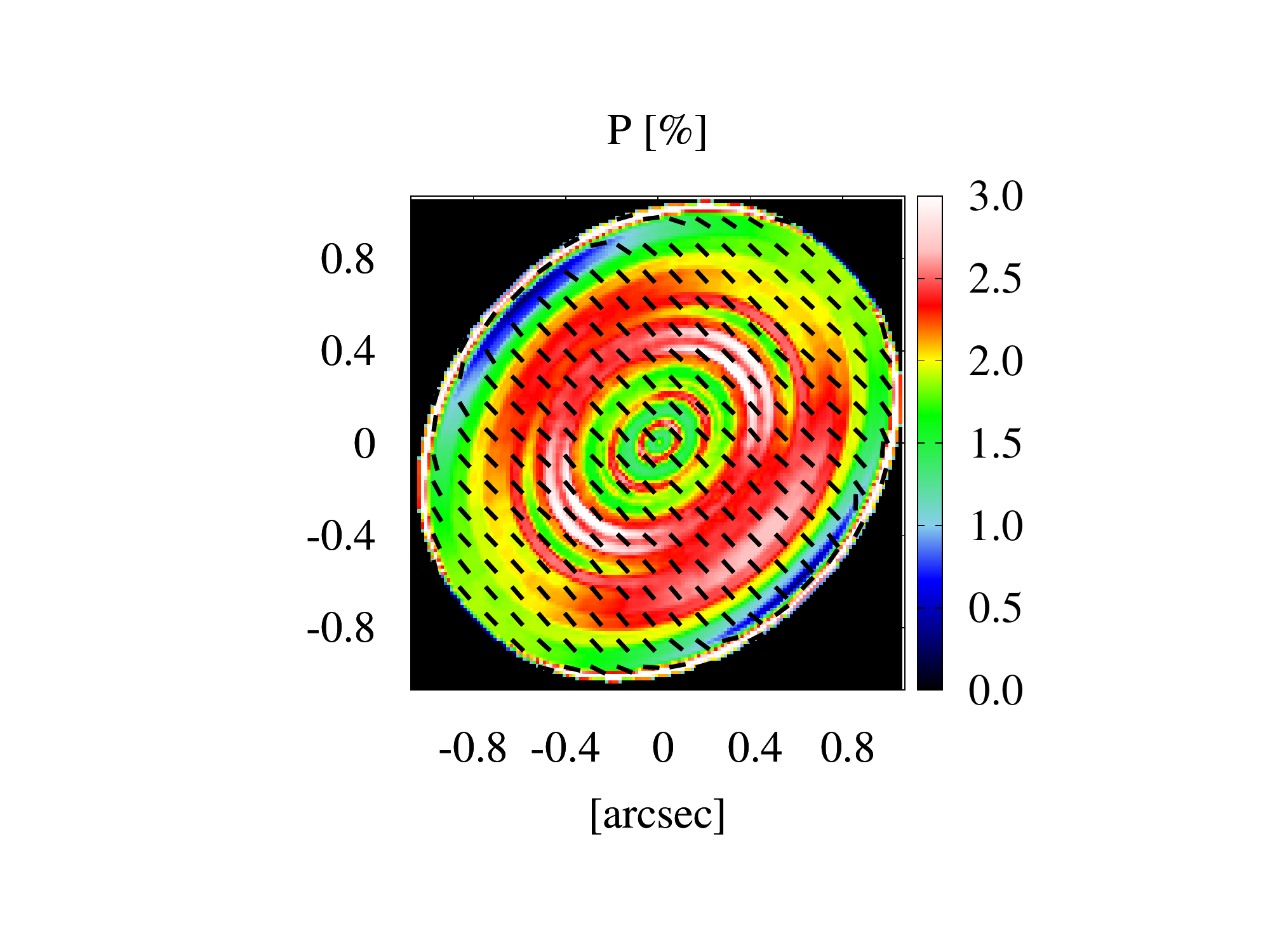}
 \hspace{-100pt}
}
\caption{
{\bf (left)} The intensity map of the radiative transfer calculations in the unit of ${\rm [Jy/arcsec^2]}$.
{\bf (center)} The polarized intensity map in the unit of ${\rm [mJy/arcsec^2]}$, overlaid with the polarization vectors.
{\bf (right)} The map of the polarization degree in the unit of \%, overlaid with the polarization vectors which is the same as the middle panel.
}
\label{fig:polarization}
 \vspace{30pt}
\end{figure*}

We show the results in the  case that the maximum grain size is $a_{\rm max} = 150 {\rm~\mu m}$ as a fiducial case. 
Figure \ref{fig:polarization} (a) shows the dust continuum, where we confirm that the continuum image well reproduces the multiple-ring structure observed with ALMA.
Figure \ref{fig:polarization} (b) shows that the polarized intensity overlaid with the polarization vectors.
Figure \ref{fig:polarization} (c) shows that the polarized degree overlaid with the polarization vectors.
 
The basic feature of the polarized intensity and the polarization vectors can be explained with the self-polarization with the anisotropy of the thermal dust emission \citep{Kataoka15}.
The thermal emission from dust grains are originally unpolarized.
The unpolarized light is scattered by other grains because of the high scattering opacity.
Due to the ring-like structure of the emission and the inclination of the disk, the distribution of the incoming flux to the scattering dust grains has anisotropies (see also \citealt{Yang15}).
As a result, the total flux has a residual polarization corresponding to the anisotropic radiation field.
Note that the scattered emission comes mainly from midplane because the disk is optically thin or marginally thick in the vertical direction (see Equation (\ref{eq:tau})).

The polarized intensity is centrally concentrated.
This infers that if we detect the polarization with a marginal sensitivity, we can detect the central part of the disk.
This is consistent with the results of polarization observation with CARMA \citep{Stephens14}, which show the centrally concentrated polarized intensity.
In addition, the polarization vectors are directed from top left to bottom right (north-east to south-west).
This is also consistent with the polarization observation with CARMA \citep{Stephens14}.
Figure \ref{fig:polarization} (c) shows the polarization degree overlaid with polarization vectors.
The polarization degree is two times higher in the gap regions than that in ring regions.
At present, the polarization image of CARMA does not have a spatial resolution high enough to resolve the rings.
Future observations of ALMA with spatial resolution as high as the long baseline campaign \citep{Partnership15} would reveal these structure even in the polarization.

\begin{figure*}[ht]
\centering
 \subfigure{
\includegraphics[width=90mm]{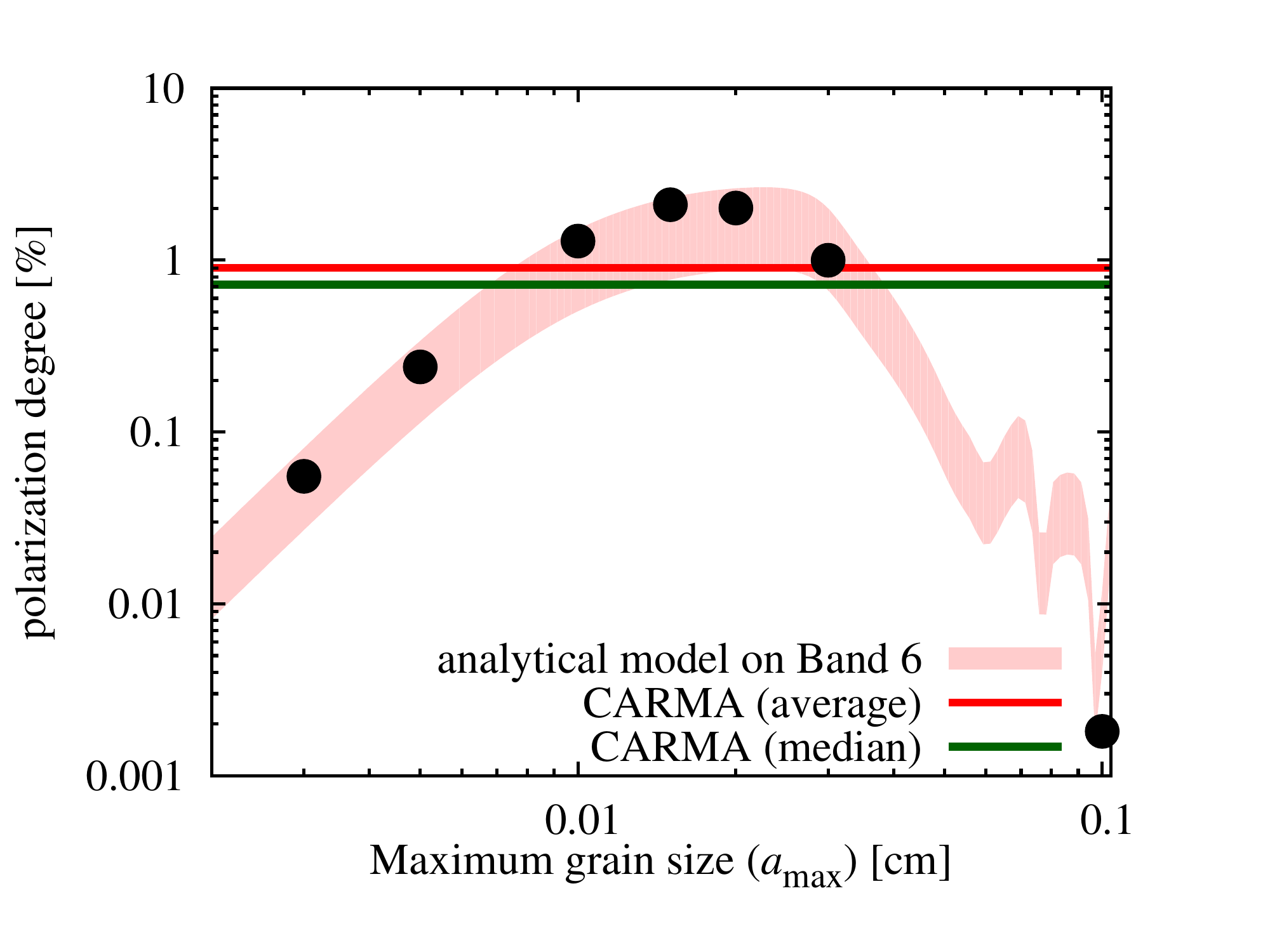}
}
 \subfigure{
\includegraphics[width=90mm ,bb=0 0 600 400]{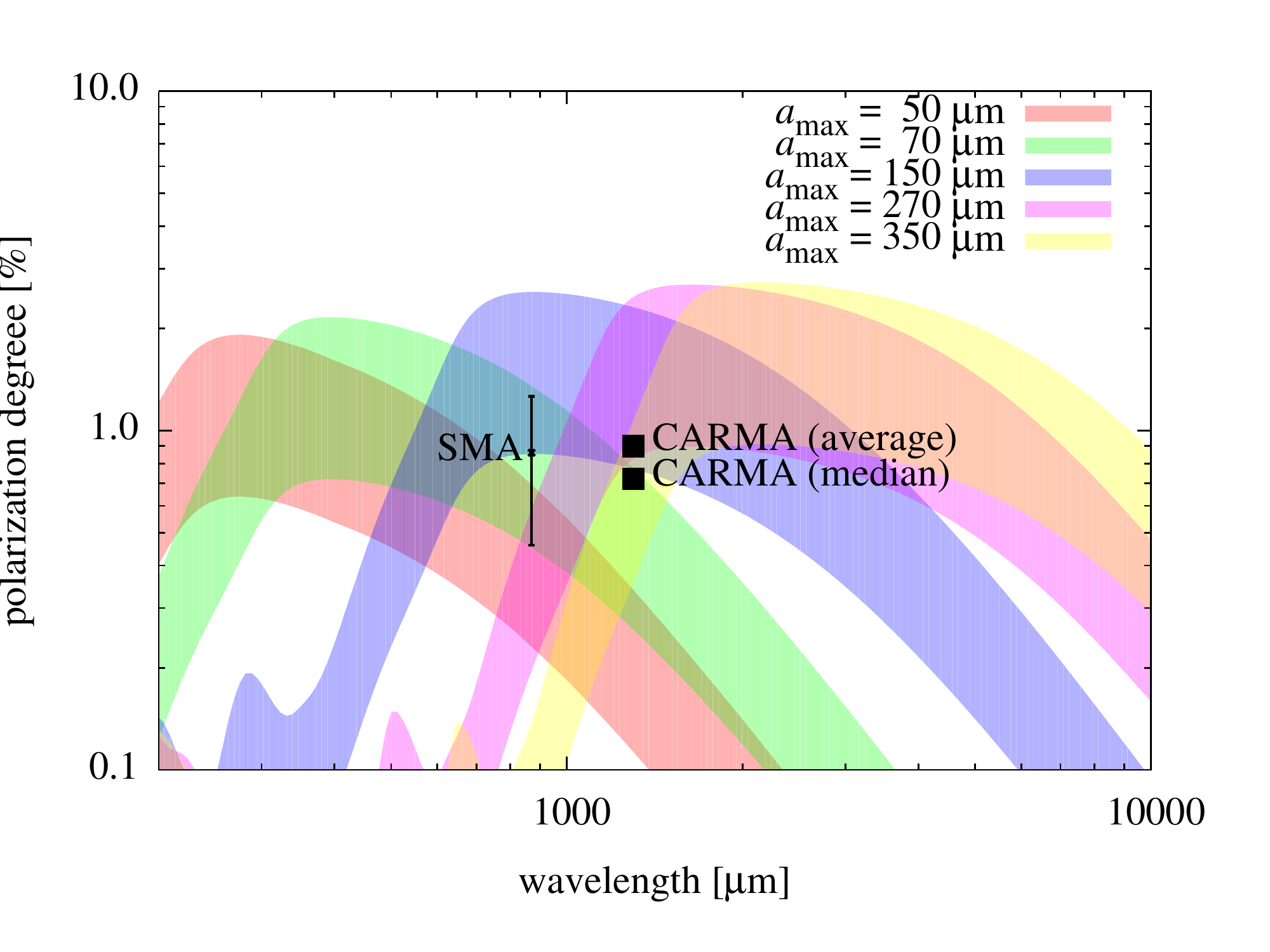}
}
\caption{
{\bf (top)}
The net polarization degrees of 0.90\% (average) and 0.72\% (median) with CARMA observations are shown as red and green lines \citep{Stephens14}.
The red shaded region is the analytical model of the net polarization degree.
The black points represent the net polarization degree of each radiative transfer calculations.
The analytical model has a good agreement with radiative transfer calculations.
From this figure, the grain size is constrained to be  $70 {\rm~\mu m} < a_{\rm max} < 350 {\rm~\mu m}$.
{\bf (bottom)}
The net polarization degrees of 0.90\% (average) and 0.72\% (median) with CARMA observations and $0.84\%\pm 0.4 \%$ with SMA observations are shown \citep{Stephens14}.
Expected polarization degree as a function of observed wavelengths.
Each thick line represents the several cases of the grain size, which are $a_{\rm max} = 50, 70, 150, 270, 350 {\rm \mu m}$.
The lines include the error of $\pm 50 \%$.
}
\label{fig:grainsize}
\end{figure*}

The reinterpretation of the polarization puts a strong constraint on the size of dust grains.
Figure \ref{fig:grainsize} shows the theoretical model of the net polarization degree at the wavelength of $\lambda=1.3 {\rm~mm}$ as a function of the maximum grain size $a_{\rm max}$.
The top of Figure \ref{fig:grainsize} shows the total polarization degree for several radiative transfer calculations at $\lambda=1.3 {\rm~mm}$ as indicated black dots overlaid with the theoretical model.
The theoretical model is calculated as the product of polarization degree at $90^\circ$ scattering $P_{90}$ and the albedo $\omega$ \citep{Kataoka15}.
The expected polarization degree is 
\begin{equation}
P=CP_{90}\omega,
\end{equation}
where $C=2.0~\%$ calibrated with the radiative transfer calculations in this paper.
We also includes the uncertainty of the model of $\pm ~50~\%$ (see Appendix of \citealt{Kataoka15} for the error of the polarization degree in different radiative transfer calculation codes).
The observed polarization degree with CARMA is also shown.
This figure clearly shows that the observed polarization degree of 0.89 \% at $\lambda=1.3 {\rm~mm}$ observed with CARMA can only be explained with the grain size in the range of $70 {\rm~\mu m} < a_{\rm max} < 350 {\rm~\mu m}$.
If the grain size is much lower, the scattering opacity is too low to scatter the thermal dust emission.
If the grain size is much higher, the scattering is forwardly peaked and thus no polarization can be expected.

We put  further constraints on the grain size with the results of SMA.
The polarization degree is a function of the combination of the observed wavelength and the grain size.
Therefore, the results of the different wavelengths put further constraints.
The bottom of Figure \ref{fig:grainsize} shows the expected polarization degree as a function of observed wavelengths with changing the maximum grain size $a_{\rm max}$.
As shown in the figure, the result of $a_{\rm max} = 350{\rm~\mu m}$ explains the CARMA observations at $\lambda=1.3{\rm~mm}$ but underestimates SMA observations at $\lambda=0.89{\rm~mm}$.
Therefore, we can rule out the possibility of $a_{\rm max} = 350{\rm~\mu m}$ to explain the polarization.
In this way, we put further constraints on the grain size to be in the range of $70 {\rm~\mu m} < a_{\rm max} < 270 {\rm~\mu m}$ to explain the polarization degree of the both CARMA and SMA observations.
We take $a_{\rm max} = 150{\rm~\mu m}$ as a representative value and continue the discussion.

\section{Discussion}

\subsection{Grain size constraints with opacity index}
The grain size has been constrained with the spectral index at sub-mm wavelengths.
If the emission is optically thin, the index has the information of the dust opacity index at sub-mm wavelengths.
The opacity index is typically equal to or less than 1 in protoplanetary disks, which can be explained with millimeter-sized grains \citep[e.g.,][]{BeckwithSargent91}.
In the case of HL Tau, the dust opacity index is in the range from 0.3 to 0.8 for the bright rings \citep{Partnership15, Kwon11}.
Therefore, the maximum grain size of 150 ${\rm \mu m}$, which is obtained in this paper, is not consistent the constraints on the grain size with the opacity index if the emission is optically thin.

Figure \ref{fig:beta} shows the opacity index $\beta$ in the case of $q=-3.5$ and $q=-2.5$ in the case of the dust model adopted in this paper.
We change the power-law index because there is no solution if the power of the grain size distribution is $q=-3.5$.
To explain the observed opacity index, the grain size should be in the range of $1 {\rm~cm} \lesssim a_{\rm max} \lesssim 5 {\rm~cm}$ if the power of the grain size distribution is $q=-2.5$. 
The opacity index strongly depends on the composition of the dust grains.
For example, dust grains composed of silicate and carbonaceous materials can produce the lower value of beta than the ice-included grains; the grains with mixture of silicate and carbonaceous material can reach $\beta=1$ even with $a_{\rm max}=500{~\mu m}$ (see Figure 4 of \citealt{Testi14} for example).
Although there are uncertainties in compositions, the maximum grain size expected from the interpretation of the spectral index is significantly larger than that expected from the polarization.

\subsection{Porous dust aggregates as a possible solution}
Here, we discuss the porous dust aggregates as a possible solution to solve the inconsistency of the grain size between the two constraints.
We have constrained the size of maximum dust grains under the assumption of spherical grains.
The upper limit of the maximum grain size is determined by the fact that the large spherical dust grains do not show the polarization due to scattering but forwardly scatter the light.
However, if we consider porous dust aggregates, the properties of their scattering and resultant polarization reflect the properties of constituent particles \citep{Min15,Tazaki15}, which are believed to have the size of (sub-)micron size.
Therefore, if we consider dust aggregates which have a even larger size than the constraint from the millimeter-wave polarization, they may explain the polarization.
This infers that the highly porous and massive aggregates may have both of the low opacity index and the high efficiency of scattering to produce the polarization.

\begin{figure*}[ht]
\centering
 \subfigure{
\includegraphics[width=80mm]{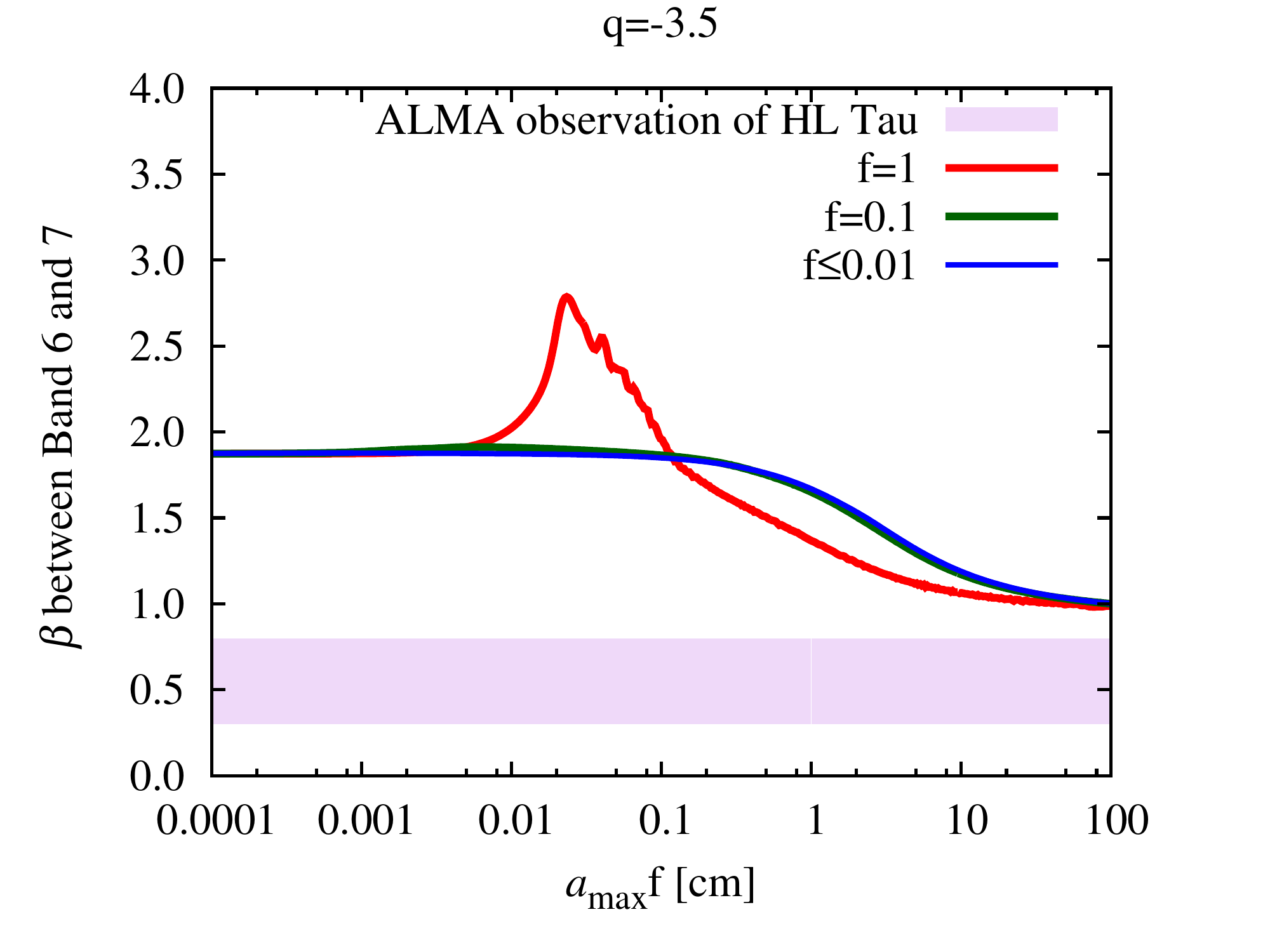}
}
 \subfigure{
\includegraphics[width=80mm]{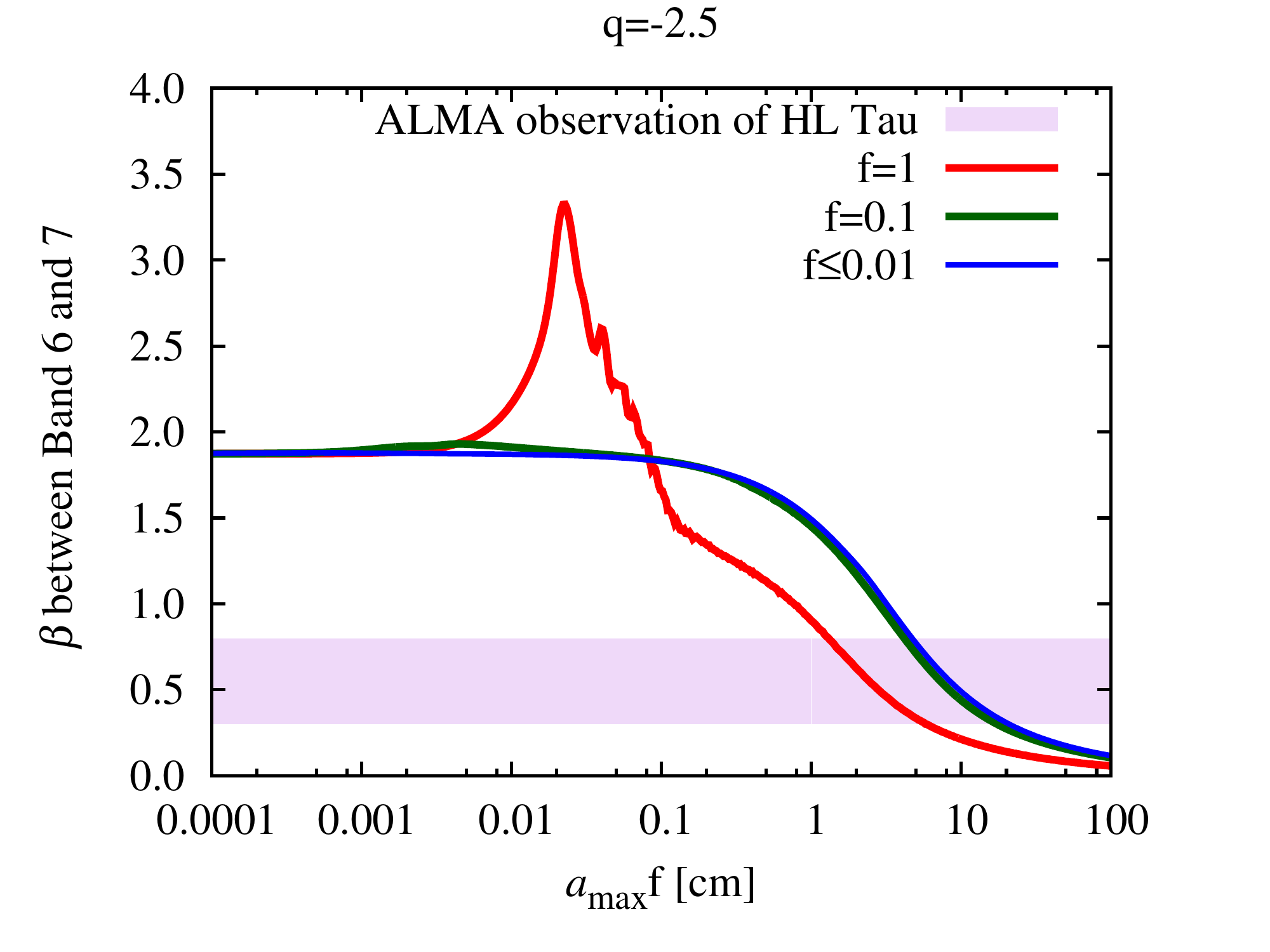}
}
\caption{
{\bf (left)} The opacity index between ALMA Band 6 (1.3 mm) and Band 7 (0.89 mm) as a function of the product of the maximum grain size $a_{\rm max}$ and the filling factor.
The power of the size distribution is taken to be $q=-3.5$.
The observed range of the opacity index is indicated as shaded region.
{\bf (right)} The power of the size distribution is taken to be $q=-2.5$.
}
\label{fig:beta}
\vspace{25pt}
\end{figure*}

To understand what kind of dust aggregates are required to explain both of the polarization and the spectral index, here we discuss the possible constraints on the dust aggregates from the spectral index assuming that the emission is optically thin.
Dust grains coagulate to form porous dust aggregates \citep[e.g.,][]{Ossenkopf93}
The filling factor can be even as small as $10^{-4}$ in disks \citep{Kataoka13b} although how porous is the dust aggregates is still controvercial.
However, we can constrain the product of the aggregate radius $a$ and the filling factor $f$ because the absorption opacity is the same if the product $af$ is the same \citep{Kataoka14}.
Figure \ref{fig:beta} also shows the opacity index as a function of the product of grain/aggregate size $a$ and the filling factor $f$.
In the case of $q=-2.5$, the grain size to explain the observations is $5{\rm~cm}\lesssim a_{\rm max}f \lesssim 21{\rm~cm}$ in the case of the fluffy aggregates ($f\le0.1$).
To explain the observed index of the absorption opacity, the dust aggregates in HL Tau should have a product of $af$ in the range of $5 {\rm~cm} \lesssim a_{\rm max}f \lesssim 21 {\rm~cm}$ in the adopted dust model.
This means that, for example, if the filling factor is as small as $f=10^{-4}$ \citep{Kataoka13b}, the aggregate radius is $\sim$ 1 km.

\begin{figure}[ht]
\centering
\includegraphics[width=\linewidth ,bb=0 0 600 400]{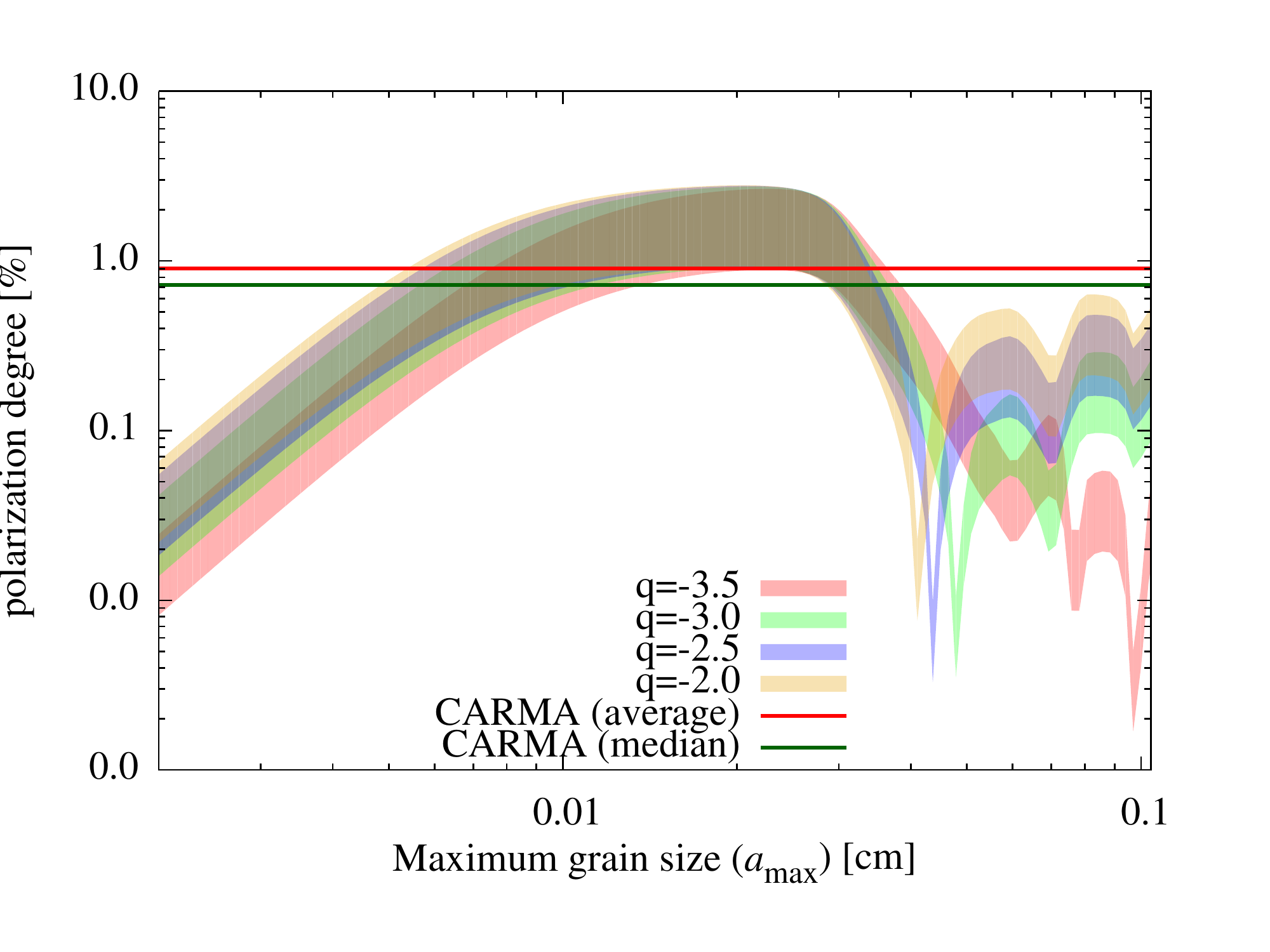}
\caption{
The expected polarization of the circumstellar disk of HL Tau with different grain size distribution.
The red and green solid lines represent CARMA observations (see Figure \ref{fig:grainsize}).
We take a power-law grain size distribution of $n(a)\propto a^{q}$ and change $q$ as $q=-3.5, -3.0, -2.5, -2.0$.
}
\label{fig:windowq}
\vspace{10pt}
\end{figure}

The size distribution does not significantly affect the polarization degree.
Figure \ref{fig:windowq} shows that the expected polarization degree in different cases of the power of the size distribution.
We do not see a significant effects on the polarization.
Therefore, we conclude that the the grain size constrained with the polarization is more robust than that with the opacity index.

\subsection{Stokes number}
The dynamics of dust grains or aggregates are determined by the normalized stopping time due to the friction between the gas and dust, which is the Stokes number.
The great advantage of knowing the product $af$ is that we can determine the Stokes number if the aggregate radius is smaller than the mean free path of the gas \citep{Kataoka14}.
If we obtain the Stokes number, we can discuss the coupling efficiency between the dust and gas.
We will discuss the Stokes number based on the discussion above.
Here, we consider two dust grain size: the 150 ${\rm \mu m}$-sized compact grains, which explain the polarization observations but do not explain the spectral index, and the 1 km-sized fluffy aggregates with $f=10^{-4}$, which explain the opacity index and which may explain the polarization properties although it is highly uncertain.
The Stokes number is written in the form of ${\rm St}\sim\rho_{\rm mat}a/\Sigma_{\rm g}$.
With the adopted model, the Stokes number is estimated to be ${\rm St}\sim 10^{-4} - 10^{-3}$ in the case of the 150 ${\rm \mu m}$ spherical grains.
If the dust aggregates are fluffy and they have $af=10{\rm~cm}$, ${\rm St}\sim 10^{-2}$.
This means that if we assume compact grains, the possible grains which can reproduce the polarization is tightly coupled to the gas. 
If we assume that fluffy dust aggregates, the coupling to the gas is relatively weak.

In the case of the spherical grains, the Stokes number is so small that they are coupled to the gas.
Therefore, if it is the case, the gas and dust have the same density distribution as the multiple-ring structure, which encourages the scenarios of gas clearance by planets \citep{Tamayo15}, magnetic instabilities of the gas \citep{Flock15}, or gravitational instability \citep{TakahashiInutsuka14}.
Furthermore, the rings may be related to enhanced grain growth at snow lines of several dust spices \citep{Zhang15}, which should be consistent with the spherical grain size because the condensation growth leads grains to be larger spherical grains.

On the other hand, if the sub-mm emission is coming from the fluffy aggregates, the Stokes number may be large enough to be decoupled from the gas.
This encourages the scenarios of trapping the dust at a radial pressure bump \citep{Pinilla12b}, at a vortex \citep{Lyra09},  or at planet-induced pressure bumps \citep{Dipierro15}.
It also support the scenario of dust fragmentation to create rings because of the higher relative velocity compared with the compact grains.
Future observations on the gas density distribution at the rings will reveal them.

In addition, the geometrical thickness of the disk also hints the grain size. 
As shown in the Appendix, the observed dust continuum image can not be reproduced without the vertical settling (see also \citealt{Pinte15}).
The vertical thickness of the disk is considerably smaller than the thermal scale height of the gas, which indicates that the dust grains are settled toward the midplane.
This requires at least the Stokes number is larger than the turbulent parameter $\alpha$ \citep[e.g.,][]{YoudinLithwick07}.
Therefore, the turbulent parameter $\alpha$ should at least be lower than $10^{-4}-10^{-3}$ in the case of compact grains and be lower than $10^{-2}$ in the case of fluffy dust aggregates from on the constraints on the Stokes number we discussed above.

\subsection{Dependence on disk models}
\label{sec:discussion}
In this paper, we have used only one disk model to constrain the grain size by the polarization signature.
However, the dependence of the polarization fraction on the disk model is weak.
As shown in Figure \ref{fig:grainsize}, the polarization fraction is described as $P=CP_{90}\omega$ and $P_{90}\omega$ is determined by the grain size \citep{Kataoka15}.
A different disk model only changes the calibration parameter $C$, which is calibrated to be 2.0\% in the adopted disk model in this paper.
Therefore, unless the disk model can change the calibration parameter $C$ in a factor of few, the constraint on the maximum grain size in this paper does not change.

\subsection{How to distinguish the polarization mechanisms?}
In this paper, we focus on the interpretation of the observed mm-wave polarization properties with the self-scattering mechanism.
However, elongated dust grains aligned with the magnetic field can also explain the polarization.
Here, we discuss the ways to distinguish the two mechanisms.
The most promising way to distinguish the polarization mechanisms is to perform the polarization observations at other wavelengths.
Although the wavelength dependence of the polarization degree is strong in the case of the self-scattering as shown in Figure \ref{fig:grainsize}, the wavelength dependence in the case of the grain alignment is not so much strong although it is still uncertain (see Figure 2 of \citealt{Andersson15} for example).
Further observations at wavelengths other than 0.87 mm and 1.3 mm will give us a clue to distinguish the two mechanisms.

\section{Conclusions}
The protoplanetary disk around HL Tau shows the polarized emission at millimeter wavelengths \citep{Tamura95, Stephens14}.
The polarized emission has been interpreted as the thermal dust emission from elongated dust grains aligned with the magnetic field.
However, the self-scattering of dust grains may also explain the observed polarization \citep{Kataoka15}.
Therefore, we have investigated whether the self-scattering of the thermal dust emission accounts for the observed millimeter-wave polarization of the protoplanetary disk around HL Tau.
We used a simple dust disk model which reasonably reproduces the millimeter-wave continuum observed with ALMA at 1.3 millimeter wavelength \citep{Partnership15}.
Dust grains are assumed to be spherical and have a power-law size distribution with the power of $q=-3.5$.
The maximum grain size $a_{\rm max}$ is the parameter.
We have performed radiative transfer calculations with the model described above with RADMC-3D to investigate the polarization properties at millimeter wavelengths.

As a result, we successfully reproduced the polarization vectors and polarization degree of HL Tau observed with CARMA and SMA\citep{Stephens14}.
We changed the maximum grain size $a_{\rm max}$ as a parameter and calculate the polarization degree to constrain the grain size.
We found that the observed polarization degree can be reproduced only if the maximum grain size is in the range of $70 {\rm~\mu m} < a_{\rm max} < 270 {\rm~\mu m}$.
This is a strong constraint on the grain size in the protoplanetary disk around HL Tau.

If the grain size is around 150 ${\rm \mu m}$, it gives a constraint on the scenario of the trapping of dust grains through to the coupling efficiency between the gas and dust.
Stokes number of these grains is estimated to be around $10^{-4}\lesssim{\rm St}\lesssim10^{-3}$, which indicates the dust grains are almost coupled to the disk gas.

We also discussed the possibility that the dust grains are porous in HL Tau disk.
The Stokes number inferred from the spectral index is as large as ${\rm St}\lesssim10^{-2}$ if the emission is optically thin and if the dust aggregates are highly porous.
If this is the case, the dust aggregates are marginally decoupled from the gas. 
However, due to the lack of knowledge about the millimeter-wave polarization of porous dust aggregates, we could not discuss the polarization degree.
The further theoretical constraints on polarization properties of porous dust aggregates are required in future studies.

We have also discussed the possible way to distinguish the mechanisms between the self-scattering and the grain alignment. 
One possible way is to perform the multi-wave observations because the polarization degree expected from the self-scattering shows strong dependence on the observed wavelength although the wavelength dependence in the case of magnetic field alignment is not so much strong.
Therefore, further observations at the wavelengths different from 1 mm is required in future studies.

\begin{acknowledgements}
This paper makes use of the following ALMA data: ADS/JAO.ALMA\#2011.0.00015.SV. 
ALMA is a partnership of ESO (representing its member states); NSF (USA); 
and NINS (Japan), together with NRC (Canada), NSC, and ASIAA (Taiwan); 
in cooperation with the Republic of Chile. The Joint ALMA Observatory 
is operated by ESO, AUI/NRAO, and NAOJ.  
This work is supported by MEXT KAKENHI No. 23103004 and by JSPS KAKENHI No. 15K17606 and 26800106

\end{acknowledgements}

\appendix
\section{Vertical settling of dust grains}
\begin{figure}[ht]
\centering
\includegraphics[width=\linewidth]{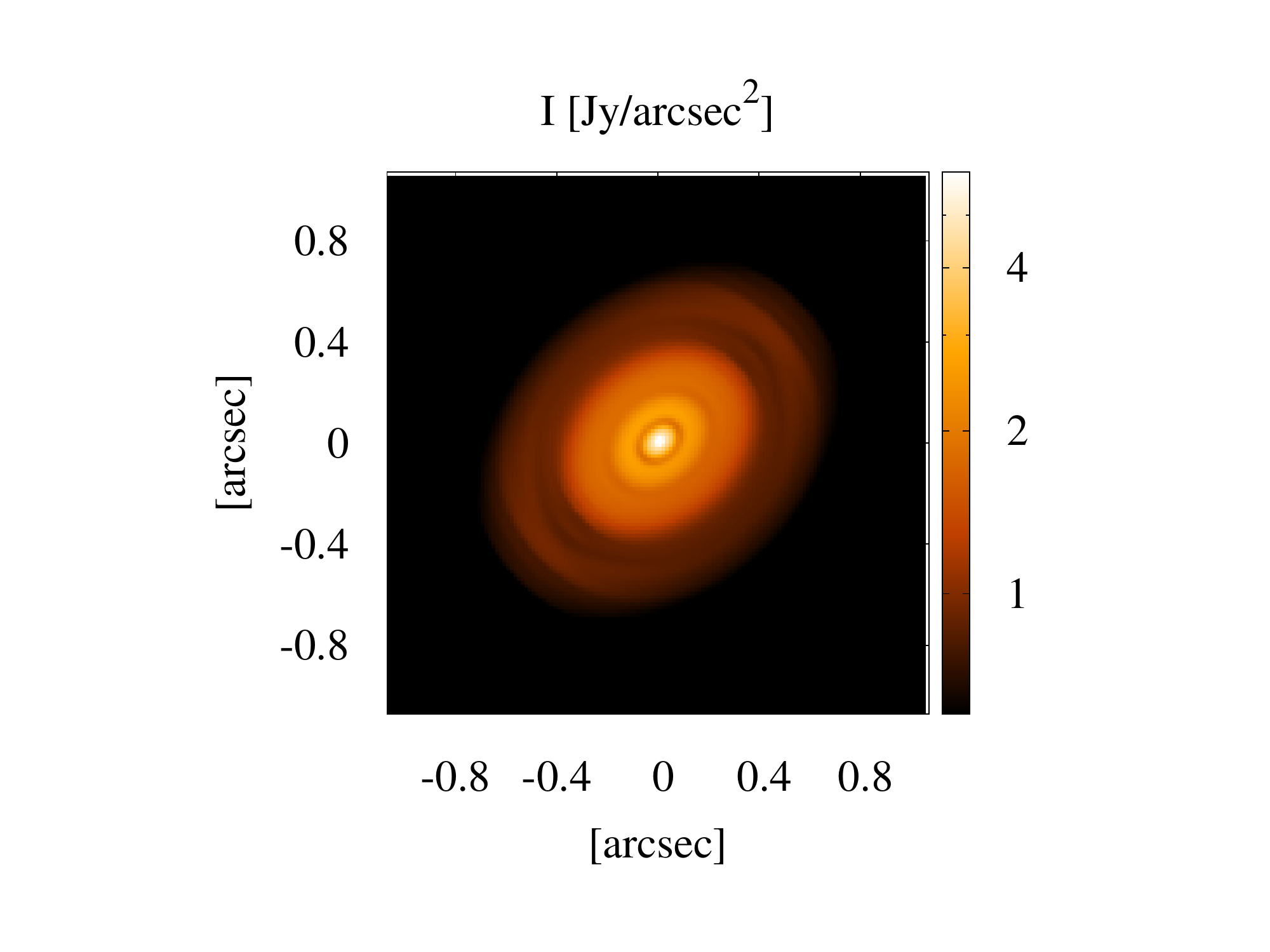}
\caption{
The intensity map in the case without the vertical settling of dust grains.
}
\label{fig:settling}
\end{figure}

We also perform the radiative transfer calculations without the dust settling; $f_{\rm settle}=1$.
In this case, the dust grains are well mixed with the gas.
The other parameters are set to be the same as the fiducial run and the maximum grain size is set to be $150 {\rm~\mu m}$.
Figure \ref{fig:settling} shows the intensity distribution.
The gaps are not clearly seen in this picture. 
This is because the scale height of dust grains are too high to reproduce the geometrically thin disk observed with ALMA (see also \citealt{Pinte15}).

\section{Dependence on disk models}
In this paper, we have used only one disk model, which reasonably reproduce the intensity distribution of ALMA observation at $\lambda=1.3{\rm~mm}$. 
Although the adopted disk model is not a unique solution, the results of the constraints on grain size would not so much depend on the disk model. 
This is because the different model of the density and temperature distribution gives different anisotropic radiation field but it only changes the absolute value of the polarization degree but does not change the relative dependence of the polarization degree on the grain size.
To demonstrate this, we perform a radiative transfer calculation of another disk model in this section.

\begin{figure}[ht]
\centering
\includegraphics[width=\linewidth]{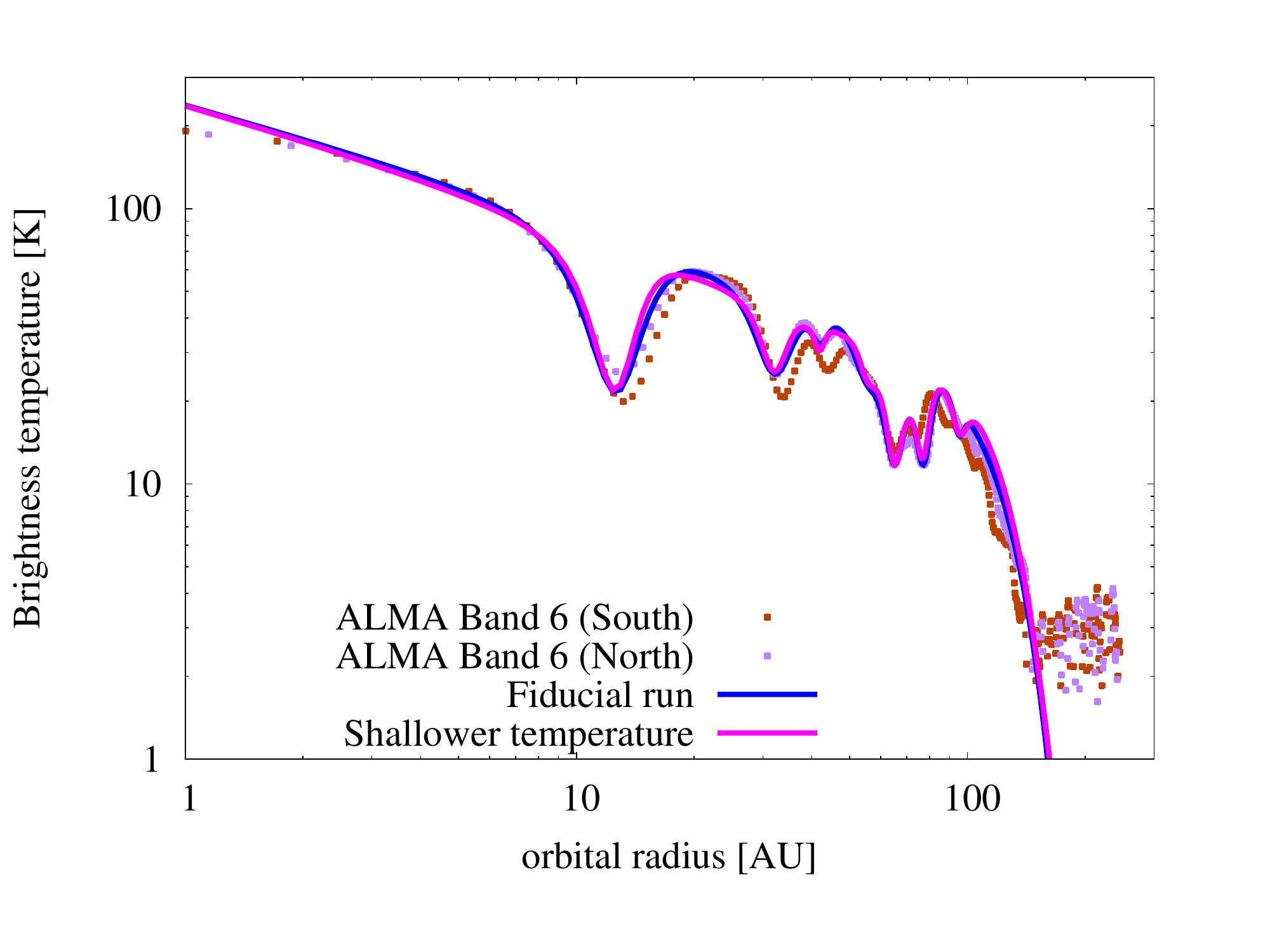}
\caption{
This figure is the same as Figure \ref{fig:model} but add the profile of the shallower temperature model adopted in the Appendix.
}
\label{fig:model_revise}
\end{figure}

\begin{figure}[ht]
\centering
\includegraphics[width=\linewidth]{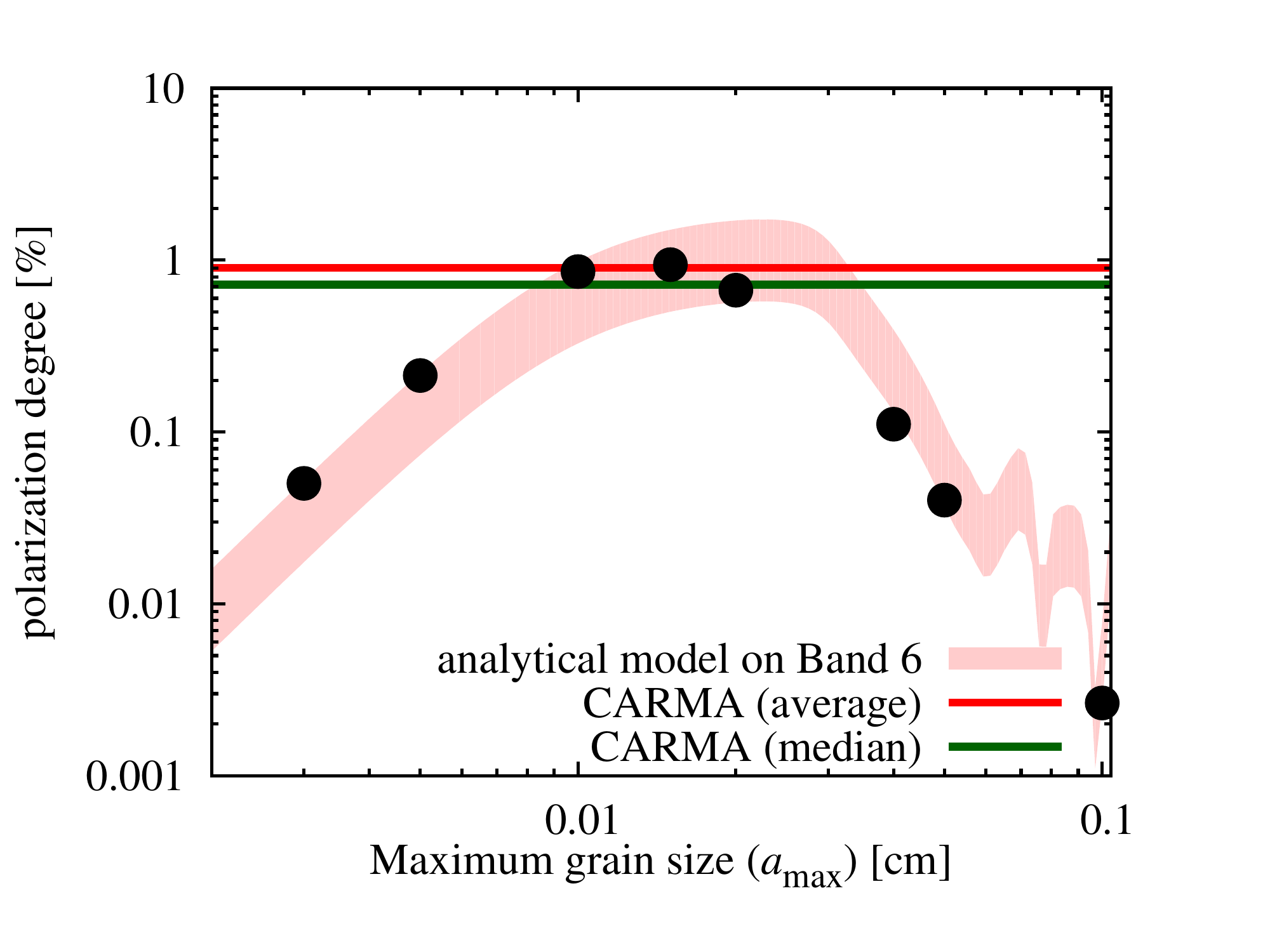}
\caption{
The same figure as Figure \ref{fig:grainsize} but for the shallower temperature slope disk model.
The analytical model has the form of $CP_{90}\omega$ where $C=1.3\%$.
}
\label{fig:grainsize_revise}
\end{figure}

We change the power-law index of temperature from $q_{\rm t}=0.3$ to $q_{\rm t}=0.5$.
In addition, to roughly fit the overall radial distribution of intensity, we set $p=-0.3$, which means that the column density increases with increasing orbital radius as $\Sigma \propto R^{0.3}$.
The parameters are summarized in Table \ref{table:gap_revise}.

Figure \ref{fig:grainsize_revise} shows the resultant polarization degree for the shallower temperature slope model.
The calibration factor $C$ of the formula $CP_{90}\omega$ is set to be $C=1.3\%$.
This figure shows that the formula $CP_{90}\omega$ can also fit the polarization fraction in the case of this model.
$P_{90}\omega$ is determined by grain size, and does not depend on the disk model.
The all contribution of the change of the disk model comes into the calibration factor $C$.
Therefore, we conclude that although the absolute value of the polarization fraction depends on the disk model, the constraints on the grain size do not so much depend on disk models; the polarization of the HL Tau can be reproduced only when the maximum grain size is around $150 {\rm~\mu m}$, although the minimum and maximum size of $a_{\rm max}$ can slightly depends on the disk model.

\begin{deluxetable}{cclcccc}
\tabletypesize{\scriptsize}
\tablecaption{The disk parameters}
\tablewidth{0pt}
\tablehead{
parameters & values
}
\startdata
$T_0$ & 280 K\\
$q_{\rm t}$ & 0.5\\
$\tau_0$&2.0\\
$p$ & -0.3\\
$s_{\rm exp}$ & 4\\
$R_{\rm exp}$ & 85 AU
\enddata
\label{table:disk_revise}
\end{deluxetable}

\begin{deluxetable}{cclcccc}
\tabletypesize{\scriptsize}
\tablecaption{The gap parameters}
\tablewidth{0pt}
\tablehead{
gap number & $r_{d,i}$ & $w_{d,i}$  & $f_{d,i}$ 
}
\startdata
1 & 12.5 & 4.0 & 1.09\\
2 & 32.0 & 5.5 & 1.2 \\
3 & 42.0 & 2.0 & 1.6 \\
4 & 55.0 & 4.0 & 1.7 \\
5 & 65.0 & 7.0 & 1.15 \\
6 & 77.0 & 5.0 & 1.3 \\
7 & 95.0 & 5.0 & 2.0 
\enddata
\label{table:gap_revise}
\end{deluxetable}

\bibliography{cite}

\end{document}